\documentclass[pre,twocolumn,superscriptaddress,showpacs]{revtex4-1}

\usepackage{graphicx}
\usepackage{dcolumn}
\usepackage{bm}
\usepackage{accents}
\usepackage{color}

\newcommand{\beq}{\begin{equation}}
\newcommand{\eeq}{\end{equation}}
\newcommand{\beqa}{\begin{eqnarray}}
\newcommand{\eeqa}{\end{eqnarray}}
\newcommand{\Tr}{\text{Tr}}

\newcommand{\rnd}[2]{\frac{\partial #1}{\partial #2}}

\newcommand{\bra}[1]{\left \langle #1 \right |}
\newcommand{\ket}[1]{\left | #1 \right \rangle}
\newcommand{\av}[1]{\left\langle #1 \right\rangle}

\newcommand{\ii}{\mathrm{i}}
\newcommand{\ee}{\mathrm{e}}

\newcommand{\var}{\text{Var}}

\newcommand{\sr}{\ln \frac{P[\Gamma]p^{\mathrm{can}}_{\lambda_{0}}(x_{0}) }{\tilde{P}[\Gamma^{\dagger}]p_{\mathrm{ini}}(x_{0})} }

\begin{document}


\title{Work fluctuation and total entropy production in nonequilibrium processes}


\author{Ken Funo}
\email[]{kenfuno@pku.edu.cn}
\affiliation{School of Physics, Peking University, Beijing 100871, China}
\author{Tomohiro Shitara}
\affiliation{Department of Physics, The University of Tokyo, 7-3-1 Hongo, Bunkyo-ku, Tokyo 113-0033, Japan}
\author{Masahito Ueda}
\affiliation{Department of Physics, The University of Tokyo, 7-3-1 Hongo, Bunkyo-ku, Tokyo 113-0033, Japan}
\affiliation{RIKEN Center for Emergent Matter Science (CEMS), 2-1 Hirosawa, Wako, Saitama 351-0198, Japan}

\date{\today}
\begin{abstract}
Work fluctuation and total entropy production play crucial roles in small thermodynamic systems subject to large thermal fluctuations. We investigate a trade-off relation between them in a nonequilibrium situation in which a system starts from an arbitrary nonequilibrium state. We apply the variational method to study this problem and find a stationary solution against  variations over protocols that describe the time dependence of the Hamiltonian of the system. Using the stationary solution, we find the minimum of the total entropy production for a given amount of work fluctuation. An explicit protocol that achieves this is constructed from an adiabatic process followed by a quasi-static process. The obtained results suggest how one can control the nonequilibrium dynamics of the system while suppressing its work fluctuation and total entropy production.
\end{abstract}


\maketitle

\section{Introduction}
Recent developments in thermodynamics of small systems based on information theoretic concepts allow one to formulate the second law of thermodynamics for arbitrary nonequilibrium initial and final states and under measurement and feedback control~\cite{Parrondo,Hasegawa,Takara,Esposito2,Deffner}. By using nonequilibrium free energies~\cite{Esposito2,Deffner}, we can quantify the extractable work from information heat engines~\cite{Maxwell,Maruyama,Sagawa2,JMaxwell,Szilard,Horowitz3,Sagawa4}, the thermodynamic cost of information erasure~\cite{Landauer,Rio} and that of a nonequilibrium thermodynamic task~\cite{Esposito2} in a unified manner. Applications of fluctuation theorems~\cite{Jarzynski1,Jarzynski2,Crooks,Esposito,fluctuation1} and stochastic thermodynamics~\cite{Sekimoto,Seifert} to these general nonequilibrium situations have been made and they provide a method to construct a protocol that reduces the entropy production during nonequilibrium processes~\cite{Hasegawa,Esposito2}.

Experimental advances, on the other hand, allow us to manipulate the microscopic degrees of freedom of small fluctuating systems. The Szilard engine and Landauer's information erasure have been demonstrated using a single-electron box~\cite{Koski1,Koski2} and a colloidal particle~\cite{Toyabe,Berut,colloidal,John}. Because the amount of work fluctuation is not negligible in mesoscopic and nano systems, much effort has been devoted to suppress excitations and work fluctuation during the nonequilibrium dynamics, thereby allowing us to obtain a faster convergence of the Jarzynski equality and an increase in the output power of heat engines~\cite{STA1,STA2,Wfluc1}. Meanwhile, the study of single-shot statistical mechanics has attracted much attention recently which applies the one-shot information theory~\cite{Renner} to thermodynamics, thereby extracting useful information on the work from a single trial of the experiment~\cite{Aberg,Horodecki,Fernando1,Fernando2,Lostaglio,Salek,Horodecki2,Oscar1,Oscar}. \color{black} In particular, a protocol with vanishing work fluctuation (deterministic work extraction protocol)~\cite{Aberg,Horodecki} and the upper bound of the unaveraged work cost (worst-case work)~\cite{Oscar1,Oscar} have been studied. The bounds on the work cost that can be derived from the deterministic work extraction protocol~\cite{Aberg,Horodecki} give severe constraints on the work compared with those of the conventional second law of thermodynamics. The basic settings to derive fluctuation theorems and the deterministic work extraction protocol are different in general, and several studies discuss the link between them~\cite{Aberg,Oscar,Salek,Funo}. In Ref.~\cite{Funo}, two of the present authors discuss a protocol which reduces both work fluctuation and total entropy production as much as possible in small thermodynamic systems. However, an implicit assumption made in that paper to derive the work fluctuation-dissipation trade-off relation turns out to be valid only for some specific range of the initial states as detailed in Appendix~\ref{sec:apendixc}. Recently, the stochastic uncertainty relation which relates the fluctuation in a current and the dissipation rate has been investigated~\cite{UC1,UC2,UC3}. Deriving some trade-off relations between fluctuation and dissipation in various nonequilibrium situations should attract considerable interest in stochastic thermodynamics.

In this paper, we remove the assumption made in Ref.~\cite{Funo} and derive a rigorous trade-off relation between the work fluctuation and the total entropy production in nonequilibrium processes for arbitrary initial states. We derive the minimum of the total entropy production for a given work fluctuation. An explicit protocol that achieves the minimum total entropy production is presented, giving us an efficient way of transforming a nonequilibrium state into a thermalized state by suppressing both work fluctuation and total entropy production as much as possible. The thermodynamically reversible protocol is reproduced in the limit of vanishing total entropy production. We derive the detailed fluctuation theorem for the single-shot setting, and show that the deterministic work extraction protocol is obtained in the limit of vanishing work fluctuation.

This paper is organized as follows. In Sec.~\ref{sec:setup}, we describe the system discussed in this paper and the assumptions made to derive the main results. In Sec.~\ref{sec:trade}, we apply the variational method to obtain the stationary solution. An explicit protocol that gives the stationary solution is given. In Sec.~\ref{sec:perturbation}, we derive the minimum of the total entropy production for a given work fluctuation by using the obtained stationary solution, which is the main result of this paper. In Sec.~\ref{sec:special}, we take the limit of vanishing total entropy production and that of vanishing work fluctuation, and show that the deterministic work extraction protocol and the thermodynamically reversible protocol are reproduced. We summarize the main results of this paper in Sec.~\ref{sec:final}. In Appendix~\ref{sec:thermal}, we derive the detailed fluctuation theorem for the dynamics of the system described by the thermal operation, and derive the deterministic work extraction protocol on the basis of the detailed fluctuation theorem. In Appendix~\ref{sec:apendixc}, we compare our main results with those obtained in Ref.~\cite{Funo}. In particular, we discuss a class of initial states such that the trade-off relation derived in Ref.~\cite{Funo} gives numerical values close to those obtained in this paper.

\section{\label{sec:setup}Setup}
We consider a situation in which the Hamiltonian of the system is externally driven according to a protocol $\lambda_{i}:=\lambda(t_{i})$ and consider discrete times $t\in\{t_{0},t_{1},\cdots,t_{N}\}$. We note that once we specify a protocol $\{\lambda_{i}\}$, the Hamiltonian of the system $H_{\lambda_{i}}$ and thus the energy eigenvalues $\{E_{\lambda_{i}}(y_{i})\}$ are specified for the entire process.  Here, we assume that the system interacts with a single heat bath whose inverse temperature is $\beta$. Suppose that the initial and final Hamiltonians ($H_{\lambda_{0}}$ and $H_{\lambda_{N}}$) are fixed, and that the initial and final states are given by $p_{\mathrm{ini}}(x_{0})$ and $p^{\mathrm{can}}_{\lambda_{N}}(x_{N})$, respectively. Here, the initial state is an arbitrary nonequilibrium distribution but the final state is assumed to be given by the canonical distribution, to disregard the nonequilibriumness of the final state which leads to a reduction in the extractable work. By this assumption, we focus on the effect of the nonequilibriumness of the initial state and discuss the optimal extractable work (which is equivalent to the minimal total entropy production as can be checked by comparing Eqs.~(\ref{defentpro}) and (\ref{wfrel}) below) for a given work fluctuation. \color{black} 
We assume that the dynamics of the system satisfies the detailed fluctuation theorem~\cite{Crooks} in Eq.~(\ref{detailedfla}). It can be derived for the classical stochastic dynamics~\cite{Seifert}, isolated quantum systems~\cite{Tasaki,Kurchan} and open quantum systems~\cite{Horowitz1,Horowitz2,Hekking,Liu1}. For quantum systems, we assume that the initial state does not have coherence between energy eigenstates. 

\subsection{\label{seccsd}Classical stochastic dynamics}
We first consider a classical Markovian dynamics described by the master equation. We consider a discrete time evolution and denote the discretized trajectory of the system as $\Gamma=\{x_{0},x_{1},\cdots,x_{N}\}$, where $x_{i}$'s denote the configuration points of the system at time $t_{i}$. Then, the system evolves in time according to the following master equation:
\beq
p(x_{i+1})=\sum_{x_{i}}p(x_{i})p(x_{i}\rightarrow x_{i+1}|H_{\lambda_{i+1}}), \label{cmaster}
\eeq
where $p(x_{i})$ is the probability of the system being found at $x_{i}$ and $p(x_{i}\rightarrow x_{i+1}|H_{\lambda_{i+1}})$ gives the transition probability of the state from $x_{i}$ to $x_{i+1}$ when the Hamiltonian of the system is given by $H_{\lambda_{i+1}}$. Note that the transition probability satisfies the following normalization condition: 
\beq
1=\sum_{x_{i+1}}p(x_{i}\rightarrow x_{i+1}|H_{\lambda_{i+1}}).\label{transition:normalize}
\eeq

We consider the following trajectory $\Gamma$ of the system and the associated change of the Hamiltonian:
\beqa
& &(H_{\lambda_{0}},x_{0})\rightarrow (H_{\lambda_{1}},x_{0})\rightarrow (H_{\lambda_{1}},x_{1}) \rightarrow (H_{\lambda_{2}},x_{1}) \nonumber \\
& &\rightarrow(H_{\lambda_{2}},x_{2}) \rightarrow \cdots \rightarrow (H_{\lambda_{N}}, x_{N-1}) \rightarrow (H_{\lambda_{N}}, x_{N}).\label{trajectorya}
\eeqa
Here, each time step is separated into the controlling substep $(H_{\lambda_{i}},x_{i})\rightarrow (H_{\lambda_{i+1}},x_{i})$ in which the Hamiltonian is changed and the relaxation substep $(H_{\lambda_{i+1}},x_{i})\rightarrow (H_{\lambda_{i+1}},x_{i+1})$ in which the state is changed by Eq.~(\ref{cmaster}). 
The forward probability distribution $P[\Gamma]$ that realizes the trajectory~(\ref{trajectorya}) is given by
\beq
P[\Gamma]:=p_{\mathrm{ini}}(x_{0})\prod_{i=0}^{N-1}p(x_{i}\rightarrow x_{i+1}|H_{\lambda_{i+1}}). \label{pforward}
\eeq
The amount of work that can be extracted from the system is defined as the energy loss of the system when its Hamiltonian is changed:
\beq
W[\Gamma]:=-\sum_{i=0}^{N-1}\left[ E_{\lambda_{i+1}}(x_{i})-E_{\lambda_{i}}(x_{i})\right]. \label{defw}
\eeq
The heat absorbed by the system is defined as the energy gain of the system via the interaction with the bath when the Hamiltonian of the system is fixed:
\beq
Q[\Gamma]:=\sum_{i=0}^{N-1}\left[ E_{\lambda_{i+1}}(x_{i+1})-E_{\lambda_{i+1}}(x_{i})\right]. \label{defq}
\eeq
Since the total energy is conserved during the relaxation process, Eq.~(\ref{defq}) is equal to the energy loss of the heat bath. Here, we note that the change in the total energy of the system can be decomposed into $W[\Gamma]$ and $Q[\Gamma]$: $E_{\lambda_{N}}(x_{N})-E_{\lambda_{0}}(x_{0})=Q[\Gamma]-W[\Gamma]$. 
The total entropy production is defined as the sum of the Shannon-entropy difference of the system $\Delta s[x_{0},x_{N}]:=\ln p_{\mathrm{ini}}(x_{0})-\ln p^{\mathrm{can}}_{\lambda_{N}}(x_{N})$ and the energy absorbed by the heat bath $-Q[\Gamma]$ multiplied by the inverse temperature $\beta$:
\beq
\sigma[\Gamma]:=\Delta s[x_{0},x_{N}]-\beta Q[\Gamma]. \label{defsigmau}
\eeq

Next, we require that the transition rates satisfy the detailed balance relation:
\beq
\frac{p(x_{i}\rightarrow x_{i+1}|H_{\lambda_{i+1}})}{p(x_{i+1}\rightarrow x_{i}|H_{\lambda_{i+1}})}=\ee^{-\beta(E_{\lambda_{i+1}}(x_{i+1})-E_{\lambda_{i+1}}(x_{i}))}. \label{localdb}
\eeq
This relation is usually assumed in stochastic thermodynamics to ensure that the system approaches thermal equilibrium when the Hamiltonian of the system is fixed~\cite{Esposito2,Seifert}. 

We now introduce the backward process  by taking the time-reversal of the protocol $\{\lambda_{i}\}$. We take the initial state of the backward process as $p^{\mathrm{can}}_{\lambda_{N}}(x_{N})$ and let the system evolve in time according to the time-reversed protocol $\tilde{\lambda}_{i}=\lambda_{N+1-i}$. We denote the time-reversed trajectory by $\Gamma^{\dagger}=\{\tilde{x}_{0},\tilde{x}_{1},\cdots,\tilde{x}_{N}\}$, where $\tilde{x}_{i}=x_{N-i}$. Then, the probability of a backward trajectory $\Gamma^{\dagger}$ being obtained is given by
\beqa
\hspace{-5mm}\tilde{P}[\Gamma^{\dagger}]&:=&p^{\mathrm{can}}_{\tilde{\lambda}_{0}}(\tilde{x}_{0})\prod_{i=0}^{N-1}p(\tilde{x}_{i}\rightarrow \tilde{x}_{i+1}|H_{\tilde{\lambda}_{i+1}}) \nonumber \\
&=&p^{\mathrm{can}}_{\lambda_{N}}(x_{N})\prod_{i=1}^{N}p(x_{N+1-i}\rightarrow x_{N-i}|H_{\lambda_{N+1-i}}). \label{pbackward}
\eeqa
If we take the ratio of the forward probability distribution to the backward probability distribution and use the  detailed balance condition~(\ref{localdb}), we obtain the detailed fluctuation theorem~\cite{Crooks}:
\beq
\frac{ P[\Gamma]  }{ \tilde{P}[\Gamma^{\dagger}]  }=\ee^{ \sigma[\Gamma]}. \label{detailedfla}
\eeq

\subsection{\label{sec:worktotal}Total entropy production and work fluctuation}
By using the detailed fluctuation theorem, the total entropy production can be expressed as the Kullback-Leibler divergence~\cite{Cover} $D(p||q):=\sum_{i}p_{i}\ln\frac{p_{i}}{q_{i}}$ between the forward process and the backward process:
\beq
\av{\sigma}=\sum_{\Gamma}P[\Gamma]\ln\frac{ P[\Gamma]}{\tilde{P}[\Gamma^{\dagger}]}=D(P||\tilde{P}) . \label{defentpro}
\eeq
The extractable work can be expressed in terms of the forward and backward probability distributions as
\beqa
W[\Gamma]&=&\beta^{-1}\biggl(\Delta s[x_{0},x_{N}]-\sigma[\Gamma]\biggr)+E_{\lambda_{0}}(x_{0})-E_{\lambda_{N}}(x_{N}) \nonumber \\
&=& -\beta^{-1}\sr +F_{\lambda_{0}}-F_{\lambda_{N}} , \label{wfrel}
\eeqa
where $ p^{\mathrm{can}}_{\lambda_{0}}(x_{0})=\exp(-\beta(E_{\lambda_{0}}(x_{0})-F_{\lambda_{0}}))$, $p^{\mathrm{can}}_{\lambda_{N}}(x_{N})=\exp(-\beta(E_{\lambda_{N}}(x_{N})-F_{\lambda_{N}}))$, and $F_{\lambda_{i}}$'s are the equilibrium free energies calculated from $H_{\lambda_{i}}$'s. By using Eq.~(\ref{wfrel}), the amount of work fluctuation is given by
\beqa
\var[W]
&=&   \beta^{-2}\Biggl[ \sum_{\Gamma}P[\Gamma]\left(\sr \right)^{2} \nonumber \\
& &\ \ \ \ \    -\left(\sum_{\Gamma}P[\Gamma] \sr \right)^{2}\Biggr]          . \label{defwfluc}
\eeqa
Here, 
we use the fact that the constant $F_{\lambda_{0}}-F_{\lambda_{N}}$ does not contribute to the variance.

\section{\label{sec:trade}Variational analysis of the relation between work fluctuation and total entropy production }
We seek for the minimum of the total entropy production $\av{\sigma}$ for a given work fluctuation
\beq
\var[W]=\Delta^{2}_{W}(=\text{const.}) \label{vara}
\eeq
by varying the protocol $\{\lambda_{i}\}$, which is equivalent to varying the intermediate energy eigenvalues $\{E_{\lambda_{i}}(y_{i})\}$. Note that the total entropy production and the work fluctuation depend on $\{E_{\lambda_{i}}(x_{i}),E_{\lambda_{i}}(x_{i-1})\}$ for the classical stochastic dynamics discussed in Sec.~\ref{seccsd}. 
\subsection{Variational method}
We use the method of Lagrange multipliers and introduce the following Lagrange function:
\beqa
S&:=&\av{\sigma}+\gamma_{1}\left(\beta^{2}\var[W]-\beta^{2}\Delta_{W}^{2}\right)\nonumber \\
&+&\gamma_{2}\left(\sum_{\Gamma}P[\Gamma] -1\right) +\gamma_{3}\left(\sum_{\Gamma}\tilde{P}[\Gamma^{\dagger}] -1\right), \label{sfunc}
\eeqa
where $\av{\sigma}$ and $\var[W]$ are given by Eqs.~(\ref{defentpro}) and (\ref{defwfluc}), respectively. 
Also, $\gamma_{1}$, $\gamma_{2}$ and $\gamma_{3}$ are the Lagrange multipliers that guarantee the following constraints:
\beqa
\rnd{S}{\gamma_{1}}&=& 0\ \ \  \Rightarrow \ \ \ \var[W]=\Delta^{2}_{W},  \label{dellambdaone}\\
\rnd{S}{\gamma_{2}}&=&0\ \ \ \Rightarrow \ \ \ \sum_{\Gamma}P[\Gamma]=1,   \label{dellambdatwo}\\
\rnd{S}{\gamma_{3}}&=&0\ \ \ \Rightarrow \ \ \ \sum_{\Gamma}\tilde{P}[\Gamma^{\dagger}]=1.   \label{dellambdathree}
\eeqa 
The stationary solutions are obtained by varying $S$ with respect to $\{E_{\lambda_{i}}(y_{i})\}$:
\beq
\frac{\delta S}{\delta E_{\lambda_{i}}(y_{i})}=0 \hspace{3mm}\text{for } \forall E_{\lambda_{i}}(y_{i}) .\label{stationaryeq}
\eeq

\subsection{\label{sec:stationary}Stationary solutions}
\begin{figure}[tbp]
\begin{center}
\includegraphics[width=.45\textwidth]{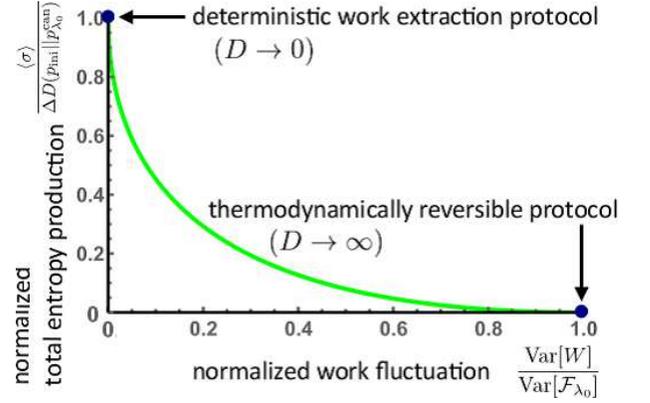}
\caption{Normalized total entropy production versus normalized work fluctuation plotted along the stationary solution~(\ref{solut}). Here, $\Delta D(p_{\mathrm{ini}}||p^{\mathrm{can}}_{\lambda_{0}}):=D(p_{\mathrm{ini}}||p^{\mathrm{can}}_{\lambda_{0}})-D_{0}(p_{\mathrm{ini}}||p^{\mathrm{can}}_{\lambda_{0}})$, with $D_{0}$ defined in Eq.~(\ref{def:renyizero}) (see also Eq.~(\ref{deterministiclimit})) and $\mathcal{F}_{\lambda_{0}}(x_{0})$ is the nonequilibrium free energy~(\ref{noneqfree}). The green curve gives the lower bound of the total entropy production for a given work fluctuation, thereby showing the boundary of the trade-off relation between work fluctuation and the total entropy production (see Sec.~\ref{sec:perturbation}). The two blue dots at both ends of the curve show the values of $(\var[W],\av{\sigma})$ obtained from the deterministic work extraction protocol~(\ref{deterministiclimit}) and the thermodynamically reversible protocol~(\ref{reversiblelimit}). }
\label{fig:i:mutual}
\end{center}
\end{figure}

As we show at the end of Sec.~\ref{sec:explicit}, Eq.~(\ref{stationaryeq}) is satisfied by the solution to the following equation:
\beq
\frac{\delta S}{\delta \tilde{P}[\Gamma^{\dagger}]}=0 \hspace{3mm}\text{for }\ \forall \Gamma. \label{sptvar}
\eeq
In the following, we obtain the stationary solution by first substituting Eq.~(\ref{sfunc}) into Eq.~(\ref{sptvar}) and obtain
\beq
\frac{1}{2\gamma_{1}}\frac{\delta S}{\delta \tilde{P}[\Gamma^{\dagger}]}=-\frac{P[\Gamma]}{\tilde{P}[\Gamma^{\dagger}]}\left[ \sr+ C\right]+D=0 , \label{barbackae}
\eeq
where we define the following two parameters
\beqa
C&:=&\frac{1}{2\gamma_{1}}-\sum_{\Gamma}P[\Gamma]\ln\frac{P[\Gamma]p^{\mathrm{can}}_{\lambda_{0}}(x_{0})}{\tilde{P}[\Gamma^{\dagger}]p_{\mathrm{ini}}(x_{0})} , \label{defC} \\
D&:=&\frac{\gamma_{3}}{2\gamma_{1}} . \label{defD}
\eeqa
We can further simplify Eq.~(\ref{barbackae}) as
\beq
D\frac{\tilde{P}[\Gamma^{\dagger}]}{P[\Gamma]}\exp\left(D\frac{\tilde{P}[\Gamma^{\dagger}]}{P[\Gamma]}\right)=De^{C}\frac{p^{\mathrm{can}}_{\lambda_{0}}(x_{0})}{p_{\mathrm{ini}}(x_{0})} \ \text{for }\forall\Gamma. \label{solutn} 
\eeq
The relevant stationary solution of our interest is the one which connects $\var[W]=0$ and $\av{\sigma}=0$ obtained from the deterministic work extraction protocol and the thermodynamically reversible protocol. It is given by
\beq
D\frac{\tilde{P}[\Gamma^{\dagger}]}{P[\Gamma]}=W_{0}\left( D\ee^{C}\frac{p^{\mathrm{can}}_{\lambda_{0}}(x_{0})}{p_{\mathrm{ini}}(x_{0})}\right) , \label{solut}
\eeq
with $0\leq D\leq \infty$ (see Fig.~\ref{fig:i:mutual}). In Eq.~(\ref{solut}), we use the upper branch of the Lambert W function ($W_{0}\geq -1$) which is defined as $W_{0}(z)=w\ \Longleftrightarrow \ z=we^{w}$~\cite{Wfunction}. 

We note that the parameters $C$ and $D$ should be determined so that the solution~(\ref{solut}) satisfies the constraints~(\ref{dellambdaone}), (\ref{dellambdatwo}) and (\ref{dellambdathree}) in the following way. We rewrite the stationary solution~(\ref{solut}) into the following form:
\beq
\tilde{P}[\Gamma^{\dagger}]=P[\Gamma|x_{0}]q(x_{0}), \label{solutqq}
\eeq
where
\beq
q(x_{0}):=\frac{1}{D}p_{\mathrm{ini}}(x_{0})W_{0}\left(De^{C} \frac{p^{\mathrm{can}}_{\lambda_{0}}(x_{0})}{p_{\mathrm{ini}}(x_{0})} \right), \label{solutzeroa}
\eeq
and we define the conditional forward probability distribution as $
P[\Gamma|x_{0}]:=P[\Gamma]/p_{\mathrm{ini}}(x_{0})$. 
By substituting the solution~(\ref{solutqq}) into the normalization condition~(\ref{dellambdathree}), we obtain
\beqa
1=  \sum_{\Gamma}\tilde{P}[\Gamma^{\dagger}]=\sum_{\Gamma}P[\Gamma|x_{0}]q(x_{0})=\sum_{x_{0}}q(x_{0}),\label{def:conda}
\eeqa
where we use Eq.~(\ref{transition:normalize}), i.e., $\sum_{x_{1},x_{2},\cdots,x_{N}}P[\Gamma|x_{0}]=1$, in deriving the last equality in Eq.~(\ref{def:conda}). From Eq.~(\ref{def:conda}), either $C$ or $D$ is fixed and from the constraint $\var[W]=\Delta^{2}_{W}$, the other parameter is determined. By determining $C$ and $D$ in this way, Eq.~(\ref{solutqq}) give the stationary solution to Eq.~(\ref{sptvar}). 


\begin{figure}[tbp]
\begin{center}
\includegraphics[width=.5\textwidth]{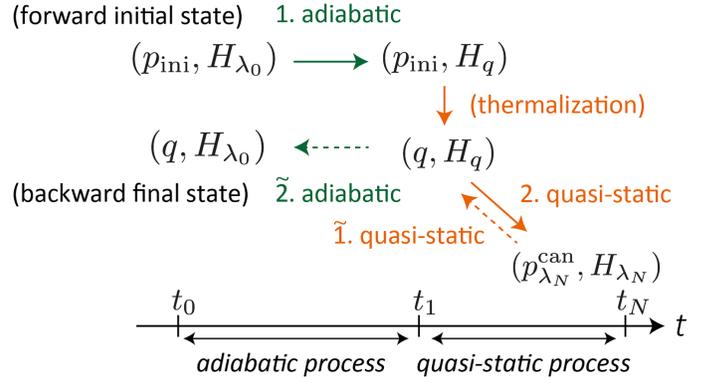}
\caption{Protocol achieving the stationary solution~(\ref{solutqq}). A change in the state of the system is shown vertically and a change in the Hamiltonian of the system is shown horizontally. The forward protocol consists of an adiabatic process followed by a quasi-static process. Here, by the quasi-static process, we mean that a change in the Hamiltonian of the system is slow compared with the relaxation of the system via the interaction with the heat bath. At the beginning of the quasi-static process, the state of the system changes from $p_{\mathrm{ini}}$ to a thermalized state $q$ given in Eq.~(\ref{solutzeroa}). In $(q,H_{q})$, $H_{q}$ represents the Hamiltonian whose canonical distribution is $q$. The backward protocol consists of a quasi-static process followed by an adiabatic process. Because the total entropy production of the forward process is nonvanishing, the final state of the backward process $q$ differs from the initial state of the forward process. }
\label{fig:i:protocol}
\end{center}
\end{figure}

\subsection{\label{sec:explicit}Explicit protocol that satisfies  the stationary condition~(\ref{sptvar})}
Now let us consider a protocol $\{\lambda_{i}\}$ that gives the stationary solution~(\ref{solutqq}). Before deriving the explicit protocol, we briefly explain the themodynamically reversible protocol for nonequilibrium initial and final states, i.e., the protocol achieving $\av{\sigma}=0$. Let us denote the forward and backward probability distributions as $P'[\Gamma]$ and $\tilde{P'}[\Gamma^{\dagger}]$, respectively. The condition $P'[\Gamma]=\tilde{P'}[\Gamma^{\dagger}]$ for $\forall \Gamma$ is satisfied if and only if the protocol is given by the thermodynamically reversible protocol. As discussed in Refs.~\cite{Esposito2,Parrondo}, this protocol can be constructed from the combination of a quench of the Hamiltonian with a quasi-static process.   

If we regard Eq.~(\ref{solutqq}) as a condition on the backward protocol, i.e., $P'[\Gamma]=\tilde{P}[\Gamma^{\dagger}]$ and $\tilde{P'}[\Gamma^{\dagger}]=P[\Gamma|x_{0}]q(x_{0})$, we find that Eq.~(\ref{solutqq}) is equivalent to the condition that the backward process is given by a thermodynamically reversible protocol that starts from $p^{\mathrm{can}}_{\lambda_{N}}(x_{N})$ and ends at $q(x_{0})$. If we denote $H_{q}$ as a Hamiltonian whose canonical distribution with the inverse temperature $\beta$ is equal to $q$, the thermodynamically reversible protocol of the backward process is given as follows (see also Fig.~\ref{fig:i:protocol}):
\begin{enumerate}
\item[$\tilde{1}$.] Quasi-statically change the Hamiltonian from $H_{\lambda_{N}}$ to $H_{q}$. Then, the state of the system changes from $p^{\mathrm{can}}_{\lambda_{N}}(x_{N})$ to $q(x_{0})$. 
\item[$\tilde{2}$.] Adiabatically change the Hamiltonian from $H_{q}$ to $H_{\lambda_{0}}$. Note that the probability distribution of the system $q(x_{0})$ does not change during this process.
\end{enumerate}
The adiabatic change of the Hamiltonian is possible if we consider either a classical system or a quantum system such that $H_{q}$ and $H_{\lambda_{0}}$ commute with each other. In this case, we can realize an adiabatic process by a sudden quench of the Hamiltonian. If $H_{q}$ and $H_{\lambda_{0}}$ do not commute with each other, we need to keep the system detached from the heat bath and change the Hamiltonian slowly so that the quantum adiabatic theorem holds.

The quasi-static process keeps the total entropy production vanishing. The adiabatic process neither changes the Shannon entropy of the system nor generates heat. The total entropy production during the above backward protocol vanishes and a thermodynamically reversible protocol for the backward process is obtained. We can also derive the forward protocol from the time-reversal of the backward protocol as follows (see also Fig.~\ref{fig:i:protocol}):
\begin{enumerate}
\item[1.] Adiabatically change the Hamiltonian from $H_{\lambda_{0}}$ to $H_{q}$. 
\item[2.] Quasi-statically change the Hamiltonian from $H_{q}$ to $H_{\lambda_{N}}$. Note that we change the Hamiltonian sufficiently slow so that the state of the system equilibrates at every step. Thus, the state of the system first changes from $p_{\mathrm{ini}}(x_{0})$ to $q(x_{0})$ via thermalization and then isothermally changes to  $p^{\mathrm{can}}_{\lambda_{N}}(x_{N})$.
\end{enumerate}

Next, let us derive explicit forms of the forward and backward probability distributions. Because an adiabatic process does not change the distribution of the system, and a quasi-static process gives the final state which is equal to the canonical distribution no matter what the initial state of the system is, the forward probability distribution is given by
\beq
P[\Gamma]=p_{\mathrm{ini}}(x_{0})p^{\mathrm{can}}_{\lambda_{N}}(x_{N}), \label{soluforward}
\eeq
and the backward probability distribution is given by 
\beq
\tilde{P}[\Gamma^{\dagger}]=p^{\mathrm{can}}_{\lambda_{N}}(x_{N})q(x_{0}). \label{solubackward}
\eeq
Here, we note that $H_{\lambda_{1}}=H_{q}$ and the quasi-static process of the backward process ends at $\tilde{t}=\tilde{t}_{N-1}(=t_{1})$ with the corresponding canonical distribution $q(x_{0})$. We also note that $E_{\lambda_{N}}(x_{N})$ is fixed. Therefore, $P[\Gamma]$ does not depend on any $\{E_{\lambda_{i}}(y_{i})\}$, and $\tilde{P}[\Gamma^{\dagger}]$ depends only on $E_{\lambda_{1}}(x_{0})=E_{q}(x_{0})$. 

From the above argument, we find that
\beq
\frac{\delta S}{\delta \tilde{P}[\Gamma^{\dagger}]}=0\ \Rightarrow\ \frac{\delta P[\Gamma]}{\delta E_{\lambda_{i}}(y_{i})}= 0\ \ \text{for }\ \forall E_{\lambda_{i}}(y_{i}). \label{deltaspi}
\eeq
By noting that $S$ is a functional of $\{E_{\lambda_{i}}(y_{i})\}$ through its dependences on $P[\Gamma]$ and $\tilde{P}[\Gamma^{\dagger}]$, Eq.~(\ref{deltaspi}) can be used to show that Eq.~(\ref{stationaryeq}) is satisfied by the stationary solutions obtained from Eq.~(\ref{sptvar}).

\section{\label{sec:perturbation}Proof of the trade-off relation between work fluctuation and the total entropy production}
In this section, we show that the stationary solution~(\ref{solutqq}) gives the global minimum of the work fluctuation for a given total entropy production. We use this stationary solution and denote the total entropy production $\av{\sigma}$ as
\beq
\Sigma:=\sum_{x_{0}}p_{\mathrm{ini}}(x_{0})\ln\frac{p_{\mathrm{ini}}(x_{0})}{q(x_{0})}, \label{global:aa} 
\eeq
and the work fluctuation $\var[\beta W]$ as
\beqa
\beta^{2}\Delta^{2}_{W}&:=&\sum_{x_{0}}p_{\mathrm{ini}}(x_{0})\left[\ln\frac{p^{\mathrm{can}}_{\lambda_{0}}(x_{0})}{q(x_{0})}\right]^{2} \nonumber \\
& &-\left(\sum_{x_{0}}p_{\mathrm{ini}}(x_{0})\ln\frac{p^{\mathrm{can}}_{\lambda_{0}}(x_{0})}{q(x_{0})}\right)^{2} . \label{global:a}
\eeqa
Note that Eqs.~(\ref{global:aa}) and (\ref{global:a}) are shown by the green solid curve in Fig.~\ref{fig:i:mutual}.

Now we consider an arbitrary protocol and denote its forward and backward probabilities as $P[\Gamma]$ and $\tilde{P}[\Gamma^{\dagger}]$, respectively. Let us divide the total entropy production of an arbitrary protocol into two parts:
\beq
\sigma[\Gamma]=\ln\frac{P[\Gamma]}{\tilde{P}[\Gamma^{\dagger}]}=\sigma_{q}(x_{0})+\Delta\sigma[\Gamma], \label{sigmace}
\eeq
where 
\beq
\sigma_{q}(x_{0})=\ln \frac{p_{\mathrm{ini}}(x_{0})}{q(x_{0})},\ \  \Delta\sigma[\Gamma]=\ln\frac{P_{q}[\Gamma]}{\tilde{P}[\Gamma^{\dagger}]}.
\label{disthermal}
\eeq
Note that $P_{q}[\Gamma]:=P[\Gamma|x_{0}]q(x_{0})$ is normalized to unity as can be seen from Eq.~(\ref{def:conda}).   
We use Eq.~(\ref{sigmace}) to calculate $\av{\sigma}$ as
\beq
\av{\sigma}=\Sigma+\av{\Delta\sigma}.\label{general:sigma}
\eeq
In what follows, we derive the global minimum of the work fluctuation for a given total entropy production $\av{\sigma}=\Sigma$. Then, we have $\av{\Delta\sigma}=0$ from Eq.~(\ref{general:sigma}), and $\var[\beta W]$ takes the form
\beqa
\hspace{-3mm}\var[\beta W]&=&\var\left[\ln\frac{p^{\mathrm{can}}_{\lambda_{0}}}{q}+\Delta\sigma\right]  \nonumber \\
&=&\beta^{2}\Delta^{2}_{W}+2\av{\Delta\sigma\ln\frac{p^{\mathrm{can}}_{\lambda_{0}}}{q}} +\av{(\Delta\sigma)^{2}},  \label{general:wf}
\eeqa
By using the relation  $\ee^{W_{0}(z)}=\frac{z}{W_{0}(z)}$, we have
\beq
2\av{\Delta\sigma\ln\frac{p^{\mathrm{can}}_{\lambda_{0}}}{q}}=2D\sum_{\Gamma}P_{q}[\Gamma]\Delta\sigma[\Gamma]\geq 0,\label{general:wo}
\eeq
where the last inequality results from $D\geq 0$ and the nonnegativity of the Kullback-Leibler divergence between $P_{q}[\Gamma]$ and $\tilde{P}[\Gamma^{\dagger}]$~\cite{Cover}:
\beq
D(P_{q}||\tilde{P})=\sum_{\Gamma}P_{q}[\Gamma]\Delta\sigma[\Gamma]\geq 0. \label{global:kl}
\eeq
Finally, we combine Eqs.~(\ref{general:wf}) and (\ref{general:wo}) and use $\av{(\Delta\sigma)^{2}}\geq 0$ to obtain
\beq
\var[W]-\Delta^{2}_{W}\geq 0.\label{global:minimum}
\eeq
We have shown that the stationary solution (\ref{solutzeroa}) gives the minimum of the work fluctuation for a given total entropy production, and the lower bound is shown by the green solid curve in Fig.~\ref{fig:i:mutual}.

We can also consider the minimum value of $\av{\sigma}$ for a given (constant) work fluctuation $\var[W]=\Delta^{2}_{W}$ in a manner similar to the derivation of Eq.~(\ref{global:minimum}). The result is equivalent to~Eq.~(\ref{global:minimum}); Eq.~(\ref{solutzeroa}) gives the minimum value of $\av{\sigma}$ for a given $\var[W]$.

The equality in~(\ref{global:minimum}) is satisfied if and only if $\Delta\sigma[\Gamma]=0$ for $\forall\Gamma$, which is equivalent to the stationary solution~(\ref{solutqq}). 
Therefore, the lower bound of the total entropy production for a given work fluctuation, depicted by the green solid curve in Fig.~\ref{fig:i:mutual}, is achieved if and only if the protocol is the one shown in Fig.~\ref{fig:i:protocol}.

\section{\label{sec:special}Some special points of the trade-off relation}
In this section, we consider some special points of the stationary solution, namely the limit of vanishing total entropy production ($D\rightarrow \infty$) and that of vanishing work fluctuation ($D\rightarrow 0$). Then, we compare those special points with the previously obtained results for the thermodynamically reversible protocol~\cite{Parrondo,Esposito2,Hasegawa,Takara} and the deterministic work extraction protocol~\cite{Aberg,Horodecki} .

\subsection{\label{sec:thermodynamically}Thermodynamically reversible protocol}
For $D=\infty$, we can use the asymptotic form of $W_{0}$ for large values of $z$: $W_{0}(z)=\ln(z)-\ln\ln(z)+\cdots$. Let us consider the normalization condition of $q(x_{0})$ by expanding $W_{0}$ up to the most divergent term: 
\beq
1=\sum_{x_{0}}q(x_{0})=\frac{C}{D}+O\left(\frac{\ln D}{D}\right) . \label{cdinfa}
\eeq
Then, using Eq.~(\ref{solutzeroa}), we obtain
\beq
q(x_{0})=p_{\mathrm{ini}}(x_{0}) \hspace{2mm} \text{ for } D=\infty. \label{reversible}
\eeq
From Eqs.~(\ref{global:aa}) and (\ref{global:a}), we find that the total entropy production vanishes; however, the amount of work fluctuation remains nonvanishing:
\beq
\av{\sigma}=0,\ \var[W]=\var[\mathcal{F}_{\lambda_{0}}], \label{reversiblelimit}
\eeq
where
\beq
\mathcal{F}_{\lambda_{0}}(x_{0}):=F_{\lambda_{0}}+kT\ln \frac{p_{\mathrm{ini}}(x_{0})}{p^{\mathrm{can}}_{\lambda_{0}}(x_{0})}\label{noneqfree}
\eeq
is the initial nonequilibrium free energy~\cite{Esposito2,Deffner}, which quantifies the maximum value of the average extractable work if the system is initially prepared in a nonequilibrium state. Note that the maximum value is achieved in this case:
\beq
\av{W}=\av{\mathcal{F}_{\lambda_{0}}}-F_{\lambda_{N}},
\eeq
and the protocol (i) and (ii) given in Sec.~\ref{sec:explicit} reproduces the thermodynamically reversible protocol discussed in Ref~\cite{Esposito}. 


\subsection{\label{sec:deterministic}Deterministic work extraction protocol}
If $D=0$, the Taylor expansion of $W_{0}$ around $0$ gives $W_{0}(z)=z-z^{2}+\cdots$. From Eq.~(\ref{solutzeroa}), we obtain 
\beq
q(x_{0})=
p^{\mathrm{can}}_{\lambda_{0}}(x_{0})\ee^{C}+O(D). \label{cdzerosoluta}
\eeq
We note that the support of $q(x_{0})$ is the same as that of $p_{\mathrm{ini}}(x_{0})$. By defining $X$ as a set of labels corresponding to the nonvanishing initial probabilities, i.e., $X=\{x|p_{\mathrm{ini}}(x_{0})>0\}$, the normalization condition~(\ref{def:conda}) determines $C=D_{0}(p_{\mathrm{ini}}||p^{\mathrm{can}}_{\lambda_{0}})$, where
\beq
D_{0}(p_{\mathrm{ini}}||p^{\mathrm{can}}_{\lambda_{0}})=-\ln \sum_{x_{0}\in X}p^{\mathrm{can}}_{\lambda_{0}}(x_{0})\label{def:renyizero}
\eeq
is the Renyi-zero divergence~\cite{Renyi}. We then obtain  
\beq
q(x_{0})
:=\biggl\{\begin{array}{lc}p^{\mathrm{can}}_{\lambda_{0}}(x_{0})\ee^{D_{0}(p_{\mathrm{ini}}||p^{\mathrm{can}}_{\lambda_{0}})}& \text{for }x_{0}\in X;\\ 0 & \text{for } x_{0}\not\in X,\end{array}  \label{qcxdzero}
\eeq
for $D=0$. Substituting Eq.~(\ref{qcxdzero}) into Eqs.~(\ref{global:aa}) and (\ref{global:a}), we have
\beq
\av{\sigma}=D(p_{\mathrm{ini}}||p^{\mathrm{can}}_{\lambda_{0}})-D_{0}(p_{\mathrm{ini}}||p^{\mathrm{can}}_{\lambda_{0}}),\ \var[W]=0. \label{deterministiclimit}
\eeq
Since the work fluctuation vanishes, the extractable work does not fluctuate and is given by
\beq
W[\Gamma]=kTD_{0}(p_{\mathrm{ini}}||p^{\mathrm{can}}_{\lambda_{0}})+F_{\lambda_{0}}-F_{\lambda_{N}}. \label{singleshotw}
\eeq

Let us define a Hamiltonian $H^{*}_{\lambda_{0}}$ which gives the canonical distribution (\ref{qcxdzero}) as follows:
\beqa
H^{*}_{\lambda_{0}}&:=&\sum_{x_{0}\in X}E_{\lambda_{0}}(x_{0})\ket{E_{\lambda_{0}}(x_{0})}\bra{E_{\lambda_{0}}(x_{0})} \nonumber \\
&+&\sum_{x_{0}\not\in X}V\ket{E_{\lambda_{0}}(x_{0})}\bra{E_{\lambda_{0}}(x_{0})},\ V\rightarrow \infty. \label{Hstar}
\eeqa
Then, $H_{q}=H^{*}_{\lambda_{0}}$ and the protocol achieving Eq.~(\ref{qcxdzero}) is given as follows: (i) Adiabatically change the Hamiltonian from $H_{\lambda_{0}}$ to $H^{*}_{\lambda_{0}}$. (ii) Quasi-statically change the Hamiltonian from $H^{*}_{\lambda_{0}}$ to $H_{\lambda_{N}}$. Note that at the beginning of (ii), the state of the system is thermalized and is given by Eq.~(\ref{qcxdzero}). Note that the $kTD_{0}(p_{\mathrm{ini}}||p^{\mathrm{can}}_{\lambda_{0}})$ term in the extractable work~(\ref{singleshotw}) is equal to the increased equilibrium free energy of the system via the adiabatic change of the Hamiltonian:
\beq
kT D_{0}(p_{\mathrm{ini}}||p^{\mathrm{can}}_{\lambda_{0}})=F^{*}_{\lambda_{0}}-F_{\lambda_{0}},\label{def:fstar}
\eeq
where $F^{*}_{\lambda_{0}}:=-\beta^{-1}\ln \text{Tr}\exp(-\beta H^{*}_{\lambda_{0}})$. The protocol (i) and (ii) reproduces the deterministic work extraction protocol discussed in Ref.~\cite{Aberg} by changing the energy levels of the system and by attaching a heat bath to the system. In Appendix~\ref{derivationdeterministic}, we consider the setups used in the single-shot statistical mechanics and reproduce the deterministic work extraction protocol discussed in Ref.~\cite{Horodecki}.

\section{\label{sec:final}Conclusion}
We have studied the minimum of the total entropy production for a given work fluctuation. By applying the variational method, we have obtained the stationary solution~(\ref{solutqq}). From the analysis performed in Sec.~\ref{sec:perturbation}, the solution~(\ref{solutqq}) is found to give the minimum of the total entropy production for a given work fluctuation in the region expressed by the green curve in Fig.~\ref{fig:i:mutual}. The protocol which achieves the minimum is shown to be constructed from an adiabatic process and a quasi-static process, as shown in Fig.~\ref{fig:i:protocol}. The obtained protocol describes an efficient way of transforming a nonequilibrium initial state to a thermalized state, thereby suppressing both work fluctuation and total entropy production. In particular, we have discussed two special ways of approaching equilibrium; one discussed in Sec.~\ref{sec:deterministic} allows the system to achieve the limit of vanishing work fluctuation, and the other discussed in Sec.~\ref{sec:thermodynamically} adiabatically transforms the system to achieve the limit of vanishing total entropy production. Below, we summarize and discuss some outstanding issues and outlooks.

We have considered the variational problem~(\ref{stationaryeq}) with respect to the protocol $\{\lambda_{i}\}$. In Sec.~\ref{sec:stationary}, we have shown that Eq.~(\ref{stationaryeq}) is satisfied by the stationary solution to the variation of the Lagrange function with the backward probability distribution. However, we have found that if we consider a variation with respect to the forward probability distribution, the stationary solution does not satisfy Eq.~(\ref{stationaryeq}). The origin of this asymmetry between the forward and backward probability distributions in the variational problem deserves further clarification. 

In Sec.~\ref{sec:explicit}, we have used the detailed fluctuation theorem and the thermodynamic reversibility of the backward protocol and obtained an explicit protocol that satisfies the stationary solutions. This method of obtaining the protocol of the system can be applied to other problems. For instance, if we place constraints on the extractable work and the total entropy production as $\alpha\sigma[\Gamma]-(1-\alpha)\beta W[\Gamma]=\text{const.}$, we obtain the protocol discussed in Ref.~\cite{Funo}, which minimizes the sum of the standard deviation of work and that of the total entropy production.

We have not considered an optimization of the protocol in a finite time, which has gathered considerable interest in recent years~\cite{Schmiedl,Aurell,Adolfo}. It is challenging to extend the obtained work-fluctuation dissipation trade-off relation to such finite-time optimization. 

We have considered the settings used in the single-shot statistical mechanics and derived the detailed fluctuation theorem in Appendix~\ref{sec:thermal}. This analysis together with the reproduced deterministic work extraction protocol from the obtained trade-off relation helps us gain deeper understanding of the relations between different approaches to thermodynamics in small systems such as the fluctuation theorems and the single-shot statistical mechanics. We note that in Ref.~\cite{Richens}, the authors investigated a connection between the deterministic work extraction protocol and the second law of thermodynamics by examining the amount of average work subject to constraints on the difference between the stochastic work and the average work.

\begin{acknowledgments}
This work was supported by KAKENHI Grant No. 26287088 from the Japan Society for the Promotion of Science, a Grant-in-Aid for Scientific Research on Innovative Areas `Topological Materials Science' (KAKENHI Grant No. 15H05855), the Photon Frontier Network Program from MEXT of Japan, and the Mitsubishi Foundation. K.F. acknowledges support from the National Science Foundation of China (grants  11375012, 11534002). T.S. acknowledges support from Grant-in-Aid for JSPS Fellows (KAKENHI Grant Number JP16J06936), and the Advanced Leading Graduate Course for Photon Science (ALPS) of JSPS. K.F. thanks Y\^{u}to Murashita for fruitful discussions and comments.
\end{acknowledgments}

\appendix

\section{\label{sec:thermal}Thermal operations and the detailed fluctuation theorem}
Usual setups in the single-shot statistical mechanics~\cite{Horodecki} are different from the setups used in stochastic thermodynamics and fluctuation theorems~\cite{Seifert}. Here we discuss how the above two setups are related so that we can compare the vanishing work fluctuation limit of the obtained trade-off relation based on the detailed fluctuation theorem with the single-shot statistical mechanics. We first derive the detailed fluctuation theorem using a  setting similar to that used in Ref.~\cite{Horodecki}, that is, the thermal operation. Then we reproduce the deterministic work extraction protocol on the basis of the obtained trade-off relation.

\subsection{Thermal operations}
In this section, we review the thermal operation which is used in Ref.~\cite{Horodecki} to derive the deterministic work extraction protocol. From the experimental point of view, the thermal operation models a state transformation of a system interacting with a single heat bath under the assumption that an arbitrary control on the system-bath coupling is possible. 

\begin{figure}[tbp]
\begin{center}
\includegraphics[width=.45\textwidth]{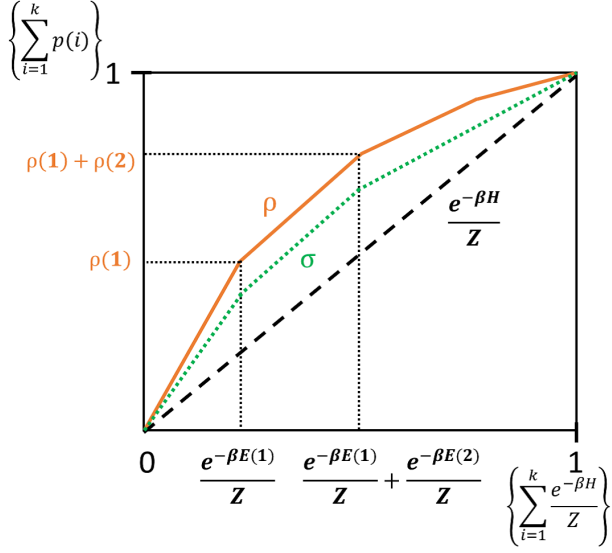}
\caption{Lorenz curve and the thermo-majorization criterion. The Lorenz curve shows a nonuniformity of the state of the system with respect to the canonical distribution, and gives a graphical representation of the quasiordering $\succ_{\mathrm{th}}$, i.e., the thermomajorization. We plot $\{\sum_{i=1}^{k}\frac{\ee^{-\beta E(i)}}{Z}\}$ and $\{\sum_{i=1}^{k}p(i)\}$ in the $(X,Y)$ plane, where the curve $p=\rho$ is shown by the orange curve, $p=\sigma$ by the green dotted curve and $p=\rho_{\mathrm{can}}=\frac{\ee^{-\beta H}}{Z}$ by the black dashed line. The thermo-majorization criterion tells us that the state $\rho$ can be transformed into $\sigma$ via a thermal operation if the Lorenz curve $(\sigma,\frac{\ee^{-\beta H}}{Z})$ is below that ot $(\rho,\frac{\ee^{-\beta H}}{Z})$. Here, the Lorenz curves are plotted for $\rho$ and $\sigma$ having the same  rearranged orderings as in Eq.~(\ref{s:dd}).}
\label{fig:s:thermal}
\end{center}
\end{figure}

Let us assume that the initial state of the system does not have coherence in the energy eigenbasis. We follow Ref.~\cite{Horodecki} and treat the external driving of the Hamiltonian as an effective dynamics of a fixed Hamiltonian of a larger system $CSW$ explained as follows. Here, the $i$-th step of the protocol changes the Hamiltonian of the system as $H_{\lambda_{i}}\rightarrow H_{\lambda_{i+1}}$ (modeled by the larger system $CSW$) followed by the relaxation of the system in contact with the heat bath $B$. We introduce a qubit system $C$ which switches the Hamiltonian of the system $S$ between $H_{\lambda_{i}}$ and $H_{\lambda_{i+1}}$ depending on the state of the qubit $\ket{0}_{C}$ or $\ket{1}_{C}$. We also introduce the work storage system $W$ which stores the work extracted from the system and define its Hamiltonian as $H_{W}:=\sum_{w}w\ket{w}\bra{w}_{W}$. Note that if we only focus on the deterministic work extraction protocol, it is enough to take a qubit system as the work storage~\cite{Horodecki}.  In the present setup, we allow fluctuations in the extracted work. Then the total Hamiltonian of the composite system $CSW$ reads
\beq
H^{\mathrm{tot}}:=\ket{0}\bra{0}_{C}\otimes H_{\lambda_{i}}+\ket{1}\bra{1}_{C}\otimes H_{\lambda_{i+1}}+H_{W}. \label{HamiltonianCSW}
\eeq
We model a general state transformation of the system due to the interaction with the heat bath by applying an arbitrary total-energy conserving unitary operator on the total system including the heat bath $B$, and take the partial trace over $B$:
\beq
\mathcal{E}_{\text{thermal}}(\rho):=\Tr_{B}\left[U\left(\rho\otimes \rho^{B}_{\mathrm{can}}\right)U^{\dagger}\right]. \label{thermaloperation}
\eeq
Here $\rho^{B}_{\mathrm{can}}=\exp(-\beta H^{B})/Z_{B}$ and $H^{B}$ are the canonical distribution and the Hamiltonian of the heat bath, respectively, and $U$ is an arbitrary unitary operator that satisfies
\beq
[U,H^{\mathrm{tot}}+H^{B}]=0. \label{thermalenergycons}
\eeq
However, $U$ is not limited to the form of $\exp(-\ii(H^{\mathrm{tot}}+H^{B})t)$ which describes the time evolution of an isolated quantum system. Implementing the thermal operation~(\ref{thermaloperation}) in an experiment is challenging because it generally requires a detailed control of the interaction between the system and the heat bath. However, from a theoretical point of view, Eq.~(\ref{thermaloperation}) can be used to search for a boundary on the allowed state transformation of the system set by thermodynamics in an extreme situation such that we have an unlimited control over the system-bath interaction. To study this boundary, thermo-majorization is introduced in Ref.~\cite{Horodecki} which establishes a quasiorder $\succ_{\mathrm{th}}$ on the density matrix of the system, giving an ``ordering'' with respect to the canonical distribution of the system. To define this quasiorder $\rho\succ_{\mathrm{th}}\sigma$, let us  denote the diagonal element of $\rho$ and that of $\sigma$ as $\rho(x)$ and $\sigma(x)$, respectively. We rearrange the label $x$ according to the following order:
\beq
\frac{\rho(1)}{\ee^{-\beta E(1)}/Z} \geq  \frac{\rho(2)}{\ee^{-\beta E(2)}/Z} \geq \frac{\rho(3)}{\ee^{-\beta E(3)}/Z} \geq \cdots. \label{s:dd}
\eeq
Then, we plot the Lorenz curve denoted as $(\frac{\ee^{-\beta H}}{Z},\rho)$ in the $(X,Y)$ plane as in Fig.~\ref{fig:s:thermal} in which each point is given by
\beqa
\thickmuskip=0mu
\medmuskip=0mu
\thinmuskip=0mu
& &\biggl\{\left(\frac{\ee^{-\beta E(1)}}{Z},\ \rho(1)\right),\left(\sum_{i=1}^{2}\frac{\ee^{-\beta E(i)}}{Z},\sum_{i=1}^{2}\rho(i)\right),\nonumber \\
& & \ \ \left(\sum_{i=1}^{3}\frac{\ee^{-\beta E(i)}}{Z},\sum_{i=1}^{3}\rho(i)\right),\ \cdots, \ (1,1)\biggr\}. \label{curve}
\eeqa
Note that the ordering~(\ref{s:dd}) ensures that the curve~(\ref{curve}) is convex. If the Lorenz curve $(\frac{\ee^{-\beta H}}{Z},\sigma)$ is below $(\frac{\ee^{-\beta H}}{Z},\rho)$, we say $\rho$ thermo-majorizes $\sigma$ and write as $\rho\succ_{\mathrm{th}}\sigma$~\cite{Majorization2}. An important property is of thermo-majorization is that $\rho\succ_{\mathrm{th}}\rho^{\mathrm{can}}$ holds for any $\rho$ and $\rho^{\mathrm{can}}:=\exp(-\beta H^{\mathrm{tot}})/Z$. Therefore, all states thermo-majorize the canonical distribution. It has been shown in~Ref.~\cite{Horodecki} that if $[\rho,H^{\mathrm{tot}}]=[\sigma,H^{\mathrm{tot}}]=0$, $\rho$ can be transformed into $\sigma$ via a thermal operation if and only if $\rho \succ_{\mathrm{th}} \sigma$. Note that the canonical distribution is a fixed point of the thermal operation, i.e., $\mathcal{E}_{\text{thermal}}(\rho^{\mathrm{can}})=\rho^{\mathrm{can}}$.

\begin{figure*}[tbp]
\begin{center}
\includegraphics[width=.95\textwidth]{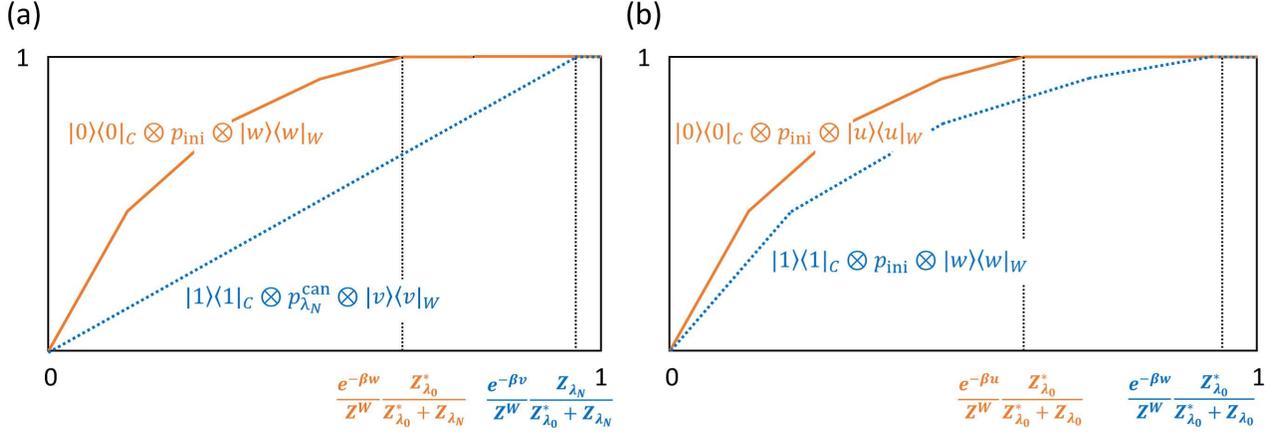}
\caption{(a) Lorenz curves corresponding to the process shown in Eq.~(\ref{thermalftr}). If $w-u\leq 0$, the blue line is below the orange curve and the transition~(\ref{thermalftr}) is possible. (b) Lorenz curves corresponding to the process shown in Eq.~(\ref{thermalftwo}). If $v-w\leq kT\ln\frac{Z_{\lambda_{N}}}{Z^{*}_{\lambda_{0}}}$, the blue curve is below the orange curve and therefore the transition~(\ref{thermalftwo}) is possible.}
\label{fig:s:thermalu}
\end{center}
\end{figure*}

\subsection{Derivation of the detailed fluctuation theorem}
We use the Hamiltonian~(\ref{HamiltonianCSW}) and consider a transition probability from 
\beq
\ket{0,x_{i},u}:=\ket{0}_{C}\otimes\ket{E_{\lambda_{i}}(x_{i})}\otimes\ket{u}_{W}
\eeq
to
\beq
\ket{1,x_{i+1},w}:=\ket{1}_{C}\otimes\ket{E_{\lambda_{i+1}}(x_{i+1})}\otimes\ket{w}_{W}.
\eeq
Note that the Hamiltonian of the system changes from $H_{\lambda_{i}}$ to $H_{\lambda_{i+1}}$ during this process. The transition probability can be calculated by using Eq.~(\ref{thermaloperation}) as
\beqa
& &p[(x_{i},u)\rightarrow (x_{i+1},w)] \nonumber \\
& &\hspace{-5mm}:=\sum_{a,b}\frac{\ee^{-\beta E^{B}_{a}}}{Z^{B}}\bigl |\bra{1,x_{i+1},w} \bra{\varphi^{B}_{b}}U\ket{\varphi^{B}_{a}}\ket{0,x_{i},u} \bigr|^{2}. \label{thermalforward}
\eeqa
Here, $\{\ket{\varphi^{B}_{a}}\}$ is the set of energy eigenvectors of the heat bath. Let us also define the backward transition probability by using the Hermitian conjugate operator of $U$:
\beqa
& &\tilde{p}[(x_{i+1},w)\rightarrow (x_{i},u)]\nonumber \\
& &\hspace{-5mm}:=\sum_{a,b}\frac{\ee^{-\beta E^{B}_{b}}}{Z^{B}} \bigl |\bra{0,x_{i},u} \bra{\varphi^{B}_{a}}U^{\dagger}\ket{\varphi^{B}_{b}}\ket{1,x_{i+1},w} \bigr|^{2}. 
\eeqa
Since thermal operations preserve the total energy [see Eq.~(\ref{thermalenergycons})], we obtain
\beq
E_{\lambda_{i}}(x_{i})+u+E^{B}_{a}=E_{\lambda_{i+1}}(x_{i+1})+w+ E^{B}_{b} \label{thermalenergyvarance}
\eeq
for the transition $\ket{\varphi^{B}_{a}}\ket{0,x_{i},u} \rightarrow \ket{1,x_{i+1},w} \ket{\varphi^{B}_{b}}$. By substituting Eq.~(\ref{thermalenergyvarance}) into Eq.~(\ref{thermalforward}) and using the relation
\beqa
\hspace{-2mm}& &\bigl |\bra{1,x_{i+1},w} \bra{\varphi^{B}_{b}}U\ket{\varphi^{B}_{a}}\ket{0,x_{i},u} \bigr|^{2} \nonumber \\
& &=\bigl |\bra{0,x_{i},u} \bra{\varphi^{B}_{a}}U^{\dagger}\ket{\varphi^{B}_{b}}\ket{1,x_{i+1},w} \bigr|^{2},
\eeqa
we obtain a relation between the forward and backward transition probabilities:
\beqa
& &p[(x_{i},u)\rightarrow (x_{i+1},w)] \nonumber \\
&=& \sum_{a,b}\biggl[ \frac{1}{Z^{B}}\ee^{-\beta (E^{B}_{b}+E_{\lambda_{i+1}}(x_{i+1})-E_{\lambda_{i}}(x_{i})+w-u)   }   \nonumber \\
& &\times\bigl |\bra{1,x_{i+1},w} \bra{\varphi^{B}_{b}}U\ket{\varphi^{B}_{a}}\ket{0,x_{i},u} \bigr|^{2} \biggr] \nonumber \\
&=& \ee^{-\beta (E_{\lambda_{i+1}}(x_{i+1})-E_{\lambda_{i}}(x_{i})+w-u) } \tilde{p}[(x_{i+1},w)\rightarrow (x_{i},u)]. \nonumber 
\eeqa
Now the heat absorbed by the system is defined as the energy decrease of the heat bath:
\beq
Q[(x_{i},u)\rightarrow (x_{i+1},w)]=-(E^{B}_{b}-E^{B}_{a}).
\eeq
It follows from Eq.~(\ref{thermalenergyvarance}) that this definition of heat is equal to the energy increase of the composite system~$CSW$:
\beq
Q[(x_{i},u)\rightarrow (x_{i+1},w)]:=E_{\lambda_{i+1}}(x_{i+1})-E_{\lambda_{i}}(x_{i})+w-u.\label{protocolthea}
\eeq
Using the definition of the heat absorbed by the system, we find that the detailed balance condition is satisfied:
\beq
\frac{p[(x_{i},u)\rightarrow (x_{i+1},w)]}{ \tilde{p}[(x_{i+1},w)\rightarrow (x_{i},u)]}=\ee^{-\beta Q[(x_{i},u)\rightarrow (x_{i+1},w)]}.
\eeq
Now, we can define the forward probability distribution as (from now on, we rewrite $u\rightarrow u_{i}$ and $w\rightarrow w_{i+1}$ for convenience)
\beq
P[\Gamma]:=p_{\mathrm{ini}}(x_{0})\prod_{i=0}^{N-1}p[(x_{i},u_{i})\rightarrow (x_{i+1},w_{i+1})],
\eeq
and the backward probability distribution as:
\beq
\thickmuskip=0mu
\medmuskip=0mu
\thinmuskip=0mu
\tilde{P}[\Gamma^{\dagger}]:=p^{\mathrm{can}}_{\lambda_{N}}(x_{N})\prod_{i=0}^{N-1}\tilde{p}[(x_{N-i},w_{N-i})\rightarrow (x_{N-i-1},u_{N-i-1})].
\eeq
We also define the total heat absorbed by the system as
\beq
Q[\Gamma]:=\sum_{i=0}^{N-1}Q[(x_{i},u_{i})\rightarrow (x_{i+1},w_{i+1})].
\eeq
We then arrive at the detailed fluctuation theorem:
\beq
\frac{ P[\Gamma]  }{ \tilde{P}[\Gamma^{\dagger}]  }=\ee^{ \sigma[\Gamma]}, \label{thermaldetailed}
\eeq
where the total entropy production is defined by Eq.~(\ref{defsigmau}). We also note that the extractable work is equal to the total excited energy of the work storage system:
\beq
W[\Gamma]=\sum_{i=0}^{N-1}(w_{i+1}-u_{i}).
\eeq

\subsection{\label{derivationdeterministic}Derivation of the deterministic work extraction protocol in Ref.~\cite{Horodecki} based on the trade-off relation}
Let us consider thermal operations and derive a protocol which realizes Eq.~(\ref{qcxdzero}). The first step is to define the total Hamiltonian as
\beq
H^{\mathrm{tot}}_{0}=H_{\lambda_{0}}\otimes\ket{0}\bra{0}_{C}+H^{*}_{\lambda_{0}}\otimes\ket{1}\bra{1}_{C}+H_{W},
\eeq
where $H^{*}_{\lambda_{0}}$ is defined in Eq.~(\ref{Hstar}). Then, we consider thermal operation that gives the following transition:
\beq
\ket{0}\bra{0}_{C}\otimes p_{\mathrm{ini}}\otimes \ket{u}\bra{u}_{W} \longrightarrow \ket{1}\bra{1}_{C}\otimes p_{\mathrm{ini}} \otimes \ket{w}\bra{w}_{W}. \label{thermalftr}
\eeq 
Here, Eq.~(\ref{thermalftr}) describes an adiabatic process that changes the Hamiltonian from $H_{\lambda_{0}}$ to $H^{*}_{\lambda_{0}}$. From the thermo-majorization curve, the transition~(\ref{thermalftr}) is possible (i.e., there exists a unitary operator $U$) if $w-u=0$ from the Lorenz curve shown in Fig.~\ref{fig:s:thermalu} (a). The next step is to combine the $N-1$ steps into one and define the total Hamiltonian as
\beq
H^{\mathrm{tot}}_{1\rightarrow N}=H^{*}_{\lambda_{0}}\otimes\ket{0}\bra{0}_{C}+H_{\lambda_{N}}\otimes\ket{1}\bra{1}_{C}+H_{W},
\eeq
and consider thermal operation that gives the following transition:
\beq
\ket{0}\bra{0}_{C}\otimes p_{\mathrm{ini}}\otimes \ket{w}\bra{w}_{W} \longrightarrow \ket{1}\bra{1}_{C}\otimes p^{\mathrm{can}}_{\lambda_{N}} \otimes \ket{v}\bra{v}_{W}. \label{thermalftwo}
\eeq 
Here, Eq.~(\ref{thermalftwo}) describes a quasi-static process that changes the Hamiltonian from $H^{*}_{\lambda_{0}}$ to $H_{\lambda_{N}}$. From the Lorenz curve shown in Fig.~\ref{fig:s:thermalu} (b), the transition~(\ref{thermalftwo}) is possible if $v-w$ is given by:
\beq
v-w=kT\ln\frac{Z_{\lambda_{N}}}{Z^{*}_{\lambda_{0}}}=F^{*}_{\lambda_{0}}-F_{\lambda_{N}}, \label{thermalfff}
\eeq
where $Z^{*}_{\lambda_{0}}=\Tr\exp(-\beta H^{*}_{\lambda_{0}})$ is the partition function. The extractable work in this case defined as the total excited energy of the work storage system and is given by
\beq
W[\Gamma]=v-u=kT D_{0}(p_{\mathrm{ini}}||p^{\mathrm{can}}_{\lambda_{0}})+F_{\lambda_{0}}-F_{\lambda_{N}}, \label{thermallimit}
\eeq
which does not fluctuate. If we combine the two thermal operations~(\ref{thermalftr}) and (\ref{thermalftwo}) into one and take $H_{\lambda_{N}}=H_{\lambda_{0}}$, we reproduce the deterministic work extraction protocol and the extractable work $kT D_{0}(p_{\mathrm{ini}}||p^{\mathrm{can}}_{\lambda_{0}})$ from a nonequilibrium system as discussed in Ref.~\cite{Horodecki}. We note that from Eq.~(\ref{thermallimit}), we only need to prepare a qubit system for the work storage $W$, whose energy difference between the excited and ground states is given by $kT D_{0}(p_{\mathrm{ini}}||p^{\mathrm{can}}_{\lambda_{0}})+F_{\lambda_{0}}-F_{\lambda_{N}}$. 

\begin{figure}[tbp]
\begin{center}
\includegraphics[width=.45\textwidth]{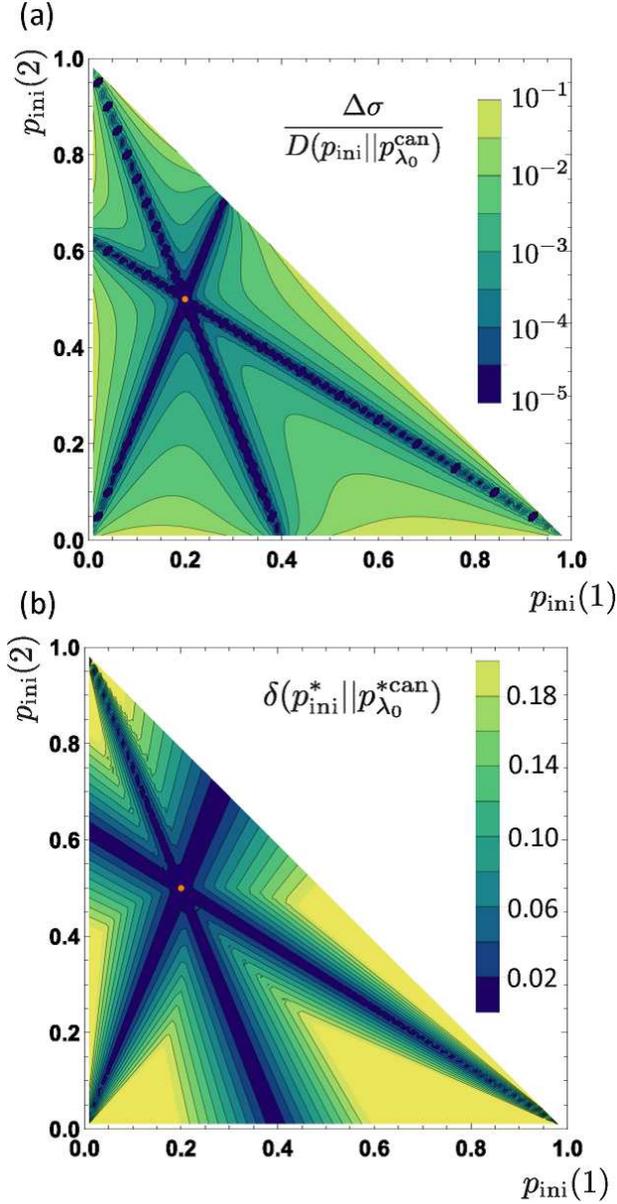}
\caption{(a) Difference between Eq.~(\ref{solutzeroa}) and Eq.~(\ref{alphasigma}) in the normalized total entropy production for different choices of the initial probability. The orange dot represents a point that satisfies $p_{\mathrm{ini}}=p^{\mathrm{can}}_{\lambda_{0}}$. We find that there is a wide region in which $p_{\mathrm{ini}}$ and $p^{\mathrm{can}}_{\lambda_{0}}$ are very different but the difference in the total entropy production is small.  (b) Plot of $\delta(p^{*}_{\mathrm{ini}}||p^{*\mathrm{can}}_{\lambda_{0}})$ for different choices of the initial probability. Here, we plot for a three-level system and fix $p^{\mathrm{can}}_{\lambda_{0}}=\{0.2,0.5,0.3\}$. }
\label{fig:i:dsigmaa}
\end{center}
\end{figure}

\section{\label{sec:apendixc}Comparison with related works}

Here, we compare the main results presented in this paper with those in Ref.~\cite{Funo}. In Ref.~\cite{Funo}, two of the present authors derived the trade-off relation between work fluctuation and dissipation by implicitly assuming that the relation $\av{f}_{\alpha}=\sum_{\Gamma}P[\Gamma]\frac{p_{\alpha}(x_{0})}{p_{\mathrm{ini}}(x_{0})}f(x_{0})$ holds even if we replace $\av{f}_{\alpha}$ by the conventional expectation values $\av{f(x_{0})}=\sum_{\Gamma}P[\Gamma]f(x_{0})$. Here, $p_{\alpha}(x)$ is defined by
\beq
p_{\alpha}(x_{0}):=[p_{\mathrm{ini}}(x_{0})]^{\alpha}[p^{\mathrm{can}}_{\lambda_{0}}(x_{0})]^{1-\alpha}\ee^{(1-\alpha)D_{\alpha}(p_{\mathrm{ini}}||p^{\mathrm{can}}_{\lambda_{0}})} \label{palphadef}
\eeq
with
\beq
D_{\alpha}(p_{\mathrm{ini}}||p^{\mathrm{can}}_{\lambda_{0}}):=\frac{1}{\alpha-1}\ln\left(\sum_{x}[p_{\mathrm{ini}}(x)]^{\alpha}[p^{\mathrm{can}}_{\lambda_{0}}(x)]^{1-\alpha}\right)
\eeq
being the Renyi divergence~\cite{Renyi}. Those two expectation values agree only when the distance between $p_{\mathrm{ini}}(x_{0})$ and $p^{\mathrm{can}}_{\lambda_{0}}(x_{0})$ is small, and thus the lower bound of the work fluctuation-dissipation trade-off relation, i.e., 
\beqa
\av{\sigma_{\alpha}}&=&D(p_{\mathrm{ini}}||p_{\alpha}), \label{alphasigma}\\
\var[W_{\alpha}]&=&\alpha^{2}\var[\mathcal{F}_{\lambda_{0}}], \label{alphawfluc}
\eeqa
derived in Ref.~\cite{Funo} does not, in general, hold for arbitrary nonequilibrium situations. However, we find that for wide choices of the initial probability distributions, the lower bound of the total entropy production for a given work fluctuation discussed in Ref.~\cite{Funo} gives numerical values close to those given by Eq.~(\ref{global:aa}) as shown in Fig.~\ref{fig:i:dsigmaa} (a). Here, we plot the difference in the total entropy production 
\beq
\Delta \sigma=\av{\sigma_{\alpha}}-\av{\sigma_{q}} \label{difentpro}
\eeq
for a constant work fluctuation by changing the initial probability distribution $p_{\mathrm{ini}}(x_{0})$ in Fig.~\ref{fig:i:dsigmaa}, (a).

\begin{figure}[t]
\begin{center}
\vspace{3mm}
\includegraphics[width=.45\textwidth]{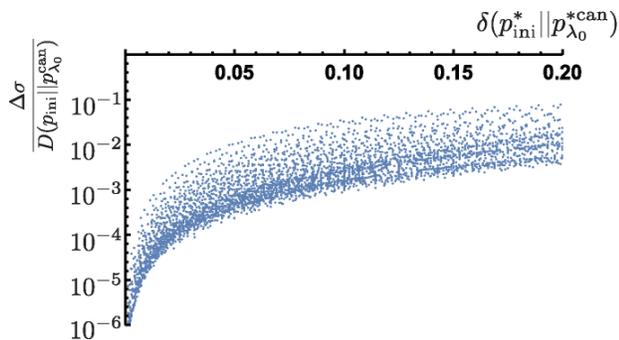}
\caption{Normalized difference in the total entropy production $\frac{\Delta\sigma}{D(p_{\mathrm{ini}}||p^{\mathrm{can}}_{\lambda_{0}})}$ versus $\delta(p^{*}_{\mathrm{ini}}||p^{*\mathrm{can}}_{\lambda_{0}})$. The data is obtained for a three-level system and $p^{\mathrm{can}}_{\lambda_{0}}=\{0.2,0.5,0.3\}$. Each dot is obtained for different choices of $p_{\mathrm{ini}}$. If the distance $\delta(p^{*}_{\mathrm{ini}}||p^{*\mathrm{can}}_{\lambda_{0}})$ is small, two lower bounds of the total entropy production in Eqs.~(\ref{global:aa}) and (\ref{alphasigma}) give approximately equal numerical values. Each dot is obtained by randomly generating $p_{\mathrm{ini}}$.}
\label{fig:i:dsigmab}
\end{center}
\end{figure}

We find from Fig.~\ref{fig:i:dsigmaa} (b) and Fig.~\ref{fig:i:dsigmab} that if the $M-1$ components of the $M$-level initial probability distribution can be approximated by the canonical distribution, we have small $\Delta \sigma$. Here, in Fig.~\ref{fig:i:dsigmab}, we plot $\Delta \sigma$ against the quantity measuring the distance between the $M-1$ components of the initial probability distribution and those of the canonical distribution:
\beqa
\hspace{-3.5mm}\delta(p^{*}_{\mathrm{ini}}||p^{*\mathrm{can}}_{\lambda_{0}})&:=&\min_{i}\delta(p^{(i)}_{\mathrm{ini}}||p^{(i),\mathrm{can}}_{\lambda_{0}})  \nonumber \\
&=&\min_{i}\frac{1}{2}\left|\sum_{x_{0}} p^{(i)}_{\mathrm{ini}}(x_{0})-p^{(i),\mathrm{can}}_{\lambda_{0}}(x_{0})\right|, \label{dstar}
\eeqa
where the ($M-1$)-level probability distributions are defined as
\beqa
p^{(i)}_{\mathrm{ini}}(x_{0})&=& \biggl\{ \frac{p_{\mathrm{ini}}(1)}{1-p_{\mathrm{ini}}(i)},\cdots,\frac{p_{\mathrm{ini}}(i-1)}{1-p_{\mathrm{ini}}(i)},\frac{p_{\mathrm{ini}}(i+1)}{1-p_{\mathrm{ini}}(i)},\nonumber \\
& &\cdots,\frac{p_{\mathrm{ini}}(M)}{1-p_{\mathrm{ini}}(i)}\biggr\}. \nonumber \\
p^{(i),\mathrm{can}}_{\lambda_{0}}(x_{0})&=&\biggl\{ \frac{p^{\mathrm{can}}_{\lambda_{0}}(1)}{1-p^{\mathrm{can}}_{\lambda_{0}}(i)},\cdots,\frac{p^{\mathrm{can}}_{\lambda_{0}}(i-1)}{1-p^{\mathrm{can}}_{\lambda_{0}}(i)},\frac{p^{\mathrm{can}}_{\lambda_{0}}(i+1)}{1-p^{\mathrm{can}}_{\lambda_{0}}(i)},\nonumber \\
& &\cdots,\frac{p^{\mathrm{can}}_{\lambda_{0}}M)}{1-p^{\mathrm{can}}_{\lambda_{0}}(i)}\biggr\}. \nonumber 
\eeqa

Note that a protocol which gives Eqs.~(\ref{alphasigma}) and (\ref{alphawfluc}) is obtained by
\beq
P[\Gamma|x_{0}]p_{\alpha}(x_{0})=\tilde{P}[\Gamma^{\dagger}]. \label{alphacond}
\eeq
If we define the following Hamiltonian
\beq
H_{\alpha}=(1-\alpha) H^{*}_{\lambda_{0}}+\alpha H_{p_{\mathrm{ini}}}, \label{alphahamiltonian}
\eeq
we find that $p_{\alpha}$ is equal to the canonical distribution with respect to $H_{\alpha}$. By comparing Eq.~(\ref{solutqq}) with Eq.~(\ref{alphacond}), we find that the protocol achieving Eqs.~(\ref{alphasigma}) and (\ref{alphawfluc}) is given by the protocol discussed in Sec.~\ref{sec:explicit} with $H_{q}$ replaced by $H_{\alpha}$. We note that $H_{\alpha}$ can be obtained by a linear combination of $H^{*}_{\lambda_{0}}$ and $H_{p_{\mathrm{ini}}}$. On the other hand, $H_{q}$ cannot be written in a simple form unlike Eq.~(\ref{alphahamiltonian}). If we consider the initial probability and the initial Hamiltonian such that $\Delta \sigma$ is sufficiently small, the protocol in Eq.~(\ref{alphacond}) is easier to implement compared with that of Eq.~(\ref{solutqq}).

\begin{thebibliography}{99}


\bibitem{Parrondo} J. M. R. Parrondo, J. M. Horowitz and T. Sagawa, {\it Thermodynamics of information}, Nat. Phys. {\bf 11}, 131 (2015).

\bibitem{Deffner} S. Deffner and E. Lutz, {\it Information free energy for nonequilibrium states}, arXiv:1201.3888.

\bibitem{Esposito2} M. Esposito and C. Van den Broeck, {\it Second law and Landauer principle far from equilibrium},  Euro. Phys. Lett. {\bf 95}, 40004 (2011).

\bibitem{Hasegawa} H.-H. Hasegawa, J. Ishikawa, K.  Takara and D. J. Driebe,  {\it Generalization of the second law for a nonequilibrium initial state}, Phys. Lett. A {\bf 374}, 1001-1004 (2010).

\bibitem{Takara} K. Takara, H.-H. Hasegawa and K. J. Driebe, {\it Generalization of the second law for a transition between nonequilibrium states}, Phys. Lett. A {\bf 375}, 88-92 (2010).

\bibitem{Maxwell} H. S. Leff and A. F. Rex, {\it Maxwell's Demon 2: Entropy, Classical and Quantum Information, Computing} (Institute of Physics Publishing, 2003).

\bibitem{Maruyama} K. Maruyama, F. Nori and V. Vedral, {\it Colloquium: The physics of Maxwell's demon and information},  Rev. Mod. Phys. {\bf 81}, 1-23 (2009).

\bibitem{Sagawa4} Thermodynamics of Information Processing in Small Systems, T. Sagawa (Springer, 2013).

\bibitem{JMaxwell} J. C. Maxwell, {\it Theory of Heat} (Appleton, London, 1871).

\bibitem{Szilard} L. Szilard, Z. Phys. {\bf 53}, 840 (1929).



\bibitem{Sagawa2} T. Sagawa and M. Ueda, {\it Second Law of Thermodynamics with Discrete Quantum Feedback Control}, Phys. Rev. Lett. {\bf 100},  080403 (2008).



\bibitem{Horowitz3} J. M. Horowitz, T. Sagawa and J. M. R. Parrondo, {\it Imitating chemical motors with optimal information motors}, Phys. Rev. Lett. {\bf 111}, 010602 (2013).


\bibitem{Landauer} R. Landauer, {\it Irreversibility and Heat Generation in the Computing Process}, IBM J. Res. Dev. {\bf 5}, 183-191 (1961).

\bibitem{Rio} L. del Rio, J. Aberg, R. Renner, O. Dahlsten and V. Vedral, Nature {\bf 474}, 61-63 (2011).










\bibitem{Jarzynski1} C. Jarzynski, {\it Nonequilibrium equality for free energy differences}, Phys. Rev. Lett. {\bf 78}, 2690 (1997).

\bibitem{Jarzynski2} C. Jarzynski, {\it Equilibrium free-energy differences from nonequilibrium measurements: A master-equation approach}, Phys. Rev. E.{\bf 56}, 5018 (1997).

\bibitem{Crooks} G. E. Crooks, {\it Entropy production fluctuation theorem and the nonequilibrium work relation for free energy differences},  Phys. Rev. E {\bf 60}, 2721-2726 (1999).

















\bibitem{Esposito} M. Esposito, U. Harbola, S. Mukamel, {\it Nonequilibrium fluctuations, fluctuation theorems, and counting statistics in quantum systems}, Rev. Mod. Phys. {\bf 81}, 1665 (2009).


\bibitem{fluctuation1} M. Campisi, P. H\"{a}nggi and P. Talkner, {\it Colloquium: Quantum fluctuation relations: Foundations and applications}, Rev. Mod. Phys. {\bf 83} 771 (2011).

\bibitem{Sekimoto} K. Sekimoto {\it Stochastic Energetics} (Lecture Notes in Physics vol 799), Springer-Verlag Berlin Heidelberg, (2010).

\bibitem{Seifert} U. Seifert, {\it Stochastic thermodynamics,  fluctuation theorems and molecular machines}, Rep. Prog. Phys. {\bf 75}, 126001 (2012).














\bibitem{Koski1} J. V. Koski, V. F. Maisi, T. Sagawa,  and J. P. Pekola, {\it Experimental Observation of the Role of Mutual Information in the Nonequilibrium Dynamics of a Maxwell Demon}, Phys. Rev. Lett. {\bf 113}, 030601 (2014).

\bibitem{Koski2} J. V. Koski, V. F. Maisi,  J. P. Pekola, and D. V. Averin, {\it Experimental realization of a Szilard engine with a single electron}, PNAS {\bf 111}, 13786 (2014).

\bibitem{Toyabe} S. Toyabe, T. Sagawa, M. Ueda, E. Muneyuki and M. Sano, {\it Experimental demonstration of information-to-energy conversion and validation of the generalized Jarzynski equality}, Nat. Phys. {\bf 6}, 988 (2010). 

\bibitem{Berut} A. B\'{e}rut, A. Arakelyan, A. Petrosyan, S. Ciliberto, R. Dillenschneider and E. Lutz, {\it Experimental verification of Landauerfs principle linking information and thermodynamics}, Nature {\bf 483}, 187-189 (2012).

\bibitem{colloidal} E. Rold\'{a}n, I. A. Martinez, J. M. R. Parrondo and D. Petrov, {\it Universal features in the energetics of symmetry breaking}, Nature Phys. {\bf 10}, 457-461 (2014).

\bibitem{John} Y. Jun, M. Gavrilov and J. Bechhoefer, {\it High-Precision Test of Landauer's Principle in a Feedback Trap}, Phys. Rev. Lett. {\bf 113}, 190601 (2014).





\bibitem{STA1} J. Deng, Q. Wang, Z. Liu, P. H\"{a}nggi and J. Gong, {\it Boosting work characteristics and overall heat-engine performance via shortcuts to adiabaticity: Quantum and classical systems}, Phys. Rev. E {\bf 88}, 062122 (2013).

\bibitem{STA2} A. del Campo, J. Goold and M. Paternostro, {\it More bang for your buck: Super-adiabatic quantum engines}, Sci. Rep. {\bf 4}, 6208 (2014).

\bibitem{Wfluc1} G. Xiao and J. Gong, {\it Suppression of work fluctuations by optimal control: An approach based on Jarzynski's equality}, Phys. Rev. E {\bf 90}, 052132 (2014).

\bibitem{Renner} R. Renner and S. Wolf, {\it Smooth Renyi entropy and applications}, ISIT p. 233 (2004).

\bibitem{Aberg} J. Aberg, {\it Truly work-like work extraction via a single-shot analysis},  Nat. Commun. {\bf 4}, 1925 (2013).

\bibitem{Horodecki} M. Horodecki and J. Oppenheim, {\it Fundamental limitations for quantum and nanoscale thermodynamics},  Nat. Commun. {\bf 4}, 2059 (2013).

\bibitem{Fernando1} F. G. S. L.Brand$\tilde{\mathrm{a}}$o, M. Horodecki, J. Oppenheim, J. M. Renes and R. W. Spekkens, {\it Resource Theory of Quantum States Out of Thermal Equilibrium}, Phys. Rev. Lett. {\bf 111}, 250404 (2013). 



\bibitem{Fernando2} F. G. S. L.Brand$\tilde{\mathrm{a}}$o, M. Horodecki, N. H. Y. Ng, J. Oppenheim and S. Wehner, {\it The second laws of quantum thermodynamics}, PNAS {\bf 112}, 3275 (2015). 

\bibitem{Lostaglio} M. Lostaglio, D. Jennings and  T. Rudolph, {\it Description of quantum coherence in thermodynamic processes requires constraints beyond free energy},  Nat. Commun. {\bf 6}, 6383 (2015).


\bibitem{Horodecki2} C. Perry, P. Cwiklinski, J. Anders, M. Horodecki and J. Oppenheim, {\it A sufficient set of experimentally implementable thermal operations}, arXiv:1511.06553.


\bibitem{Oscar1} H. Y. Halpern, A. J. Garner, O. C. Dahlsten and V. Vedral, {\it Introducing one-shot work into fluctuation relations}, New. J. Phys. {\bf 17}, 095003 (2015).

\bibitem{Salek} S. Salek and K. Wiesner, {\it Fluctuations in Single-Shot $\epsilon$-Deterministic Work Extraction}, arXiv:1504.05111.

\bibitem{Oscar} O. C. O. Dahlsten, M-S. Choi, D. Braun, A. J. P. Garner, N. Y. Halpern and V. Vedral, {\it Equality for worst-case work at any protocol speed}, arXiv:1504.05152.

\bibitem{Funo} K. Funo and M. Ueda, Phys. Rev. Lett. {\it Work Fluctuation-Dissipation Trade-Off in Heat Engines}, {\bf 115}, 260601 (2015).

\bibitem{UC1} A. C. Barato and U. Seifert, {\it Thermodynamic uncertainty relation for biomolecular processes}, Phys. Rev. Lett. {\bf 114}, 158101 (2015).

\bibitem{UC2} T. R. Gingrich, J. M. Horowitz, N. Perunov and J. England, {\it Dissipation bounds all steady-state current fluctuations}, Phys. Rev. Lett. {\bf 116}, 120601 (2016). 


\bibitem{UC3} M. Polettini, A. Lazarescu and M. Esposito, {\it Tightening the uncertainty principle for stochastic currents}, arXiv:1605.09692.


\bibitem{Tasaki} H. Tasaki, {\it Jarzynski Relations for Quantum Systems and Some Applications}, arXiv:cond-mat/0009244.

\bibitem{Kurchan} J. Kurchan, {\it A Quantum Fluctuation Theorem}, cond-mat/0007360.


\bibitem{Horowitz1} J. M. Horowitz, {\it Quantum-trajectory approach to the stochastic thermodynamics of a forced harmonic oscillator}, Phys. Rev. E {\bf 85}, 031110 (2012).

\bibitem{Horowitz2} J. M. Horowitz and J. M. R. Parrondo, {\it Entropy production along nonequilibrium quantum jump trajectories}, New J. Phys. {\bf 15} 085028 (2013).

\bibitem{Hekking} F. W. J. Hekking and J. P. Pekola, {\it Quantum Jump Approach for Work and Dissipation in a Two-Level System}, Phys. Rev. Lett. {\bf 111}, 093602 (2013).

\bibitem{Liu1} F. Liu, Phys. Rev. E {\bf 90}, {\it Calculating work in adiabatic two-level quantum Markovian master equations: A characteristic function method}, 032121 (2014).







\bibitem{Majorization2} G. Gour, M. P. Muller, V. Narasimhachar, R. W. Spekkens and N. Y. Halpern, {\it The resource theory of informational nonequilibrium in thermodynamics}, Phys. Rep. {\bf 583}, 1-58 (2015).


\bibitem{Cover} T. M. Cover and J. A. Thomas {\it Elements of information theory} (John Wiley \& Sons, 2012).

\bibitem{Wfunction} R. M. Corless, G. H. Gonnet, D. E. G. Hare, D. J. Jeffrey and D. E. Knuth, {\it On the Lambert $W$ function}, Adv. Comput. Math. {\bf 5} 329-359 (1996).


\bibitem{Renyi} A. R\'{e}nyi, {\it On measures of entropy and information}, in Proc. 4th Berkeley Symp. Math. Statist. and Probability, {\bf 1}, 547-561 (1961).

\bibitem{Schmiedl} T. Schmiedl and U. Seifert, {\it Optimal Finite-Time Processes In Stochastic Thermodynamics}, Phys. Rev. Lett. {\bf 98}, 108301 (2007).

\bibitem{Aurell} E. Aurell, C. M.-Monasterio and P. M.-Ginanneschi, {\it Optimal Protocols and Optimal Transport in Stochastic Thermodynamics}, Phys. Rev. Lett. {\bf 106}, 250601 (2011).

\bibitem{Adolfo} E. Torrontegui, S. Ib\'{a}\~{n}ez, S. Mart\'{i}nez-Garaot, M. Modugno, A. del Campo, D. Gu\'{e}ry-Odelin, A. Ruschhaupt, X. Chen, J. G. Muga, {\it Shortcuts to Adiabaticity}, Adv. At. Mol. Opt. Phys. {\bf 62}, 117-169 (2013). 

\bibitem{Richens} J. G. Richtens and L. Masanes, {\it Quantum thermodynamics with constrained fluctuations in work}, arXiv:1603.02417.





\end{thebibliography}
\end{document}